# Utility-based Decision-making in Distributed Systems Modelling

## [Extended Abstract]

Gabrielle Anderson[*]
University of Aberdeen
Scotland, UK

Matthew Collinson[†]
University of Aberdeen
Scotland, UK

David Pym[‡]
University of Aberdeen
Scotland, UK

## ABSTRACT

We consider a calculus of resources and processes as a basis for modelling decision-making in multi-agent systems. The calculus represents the regulation of agents' choices using utility functions that take account of context. Associated with the calculus is a (Hennessy–Milner-style) context-sensitive modal logic of state. As an application, we show how a notion of 'trust domain' can be defined for multi-agent systems.

## 1. INTRODUCTION

Mathematical modelling is a key tool in designing and reasoning about the complex systems of systems upon which the world depends. For modelling complex information processing systems, including both logical and physical components, the classical theory of distributed systems — see, for example, [13] for an elegant account — provides a suitable conceptual basis [11] for a modelling discipline. Executable modelling languages are important supporting tools, providing methods for simulating — using both Monte Carlo and what-if methods — systems that are too complex for useful analytical solvable descriptions. Techniques such as model checking can be applied in sufficiently constrained circumstances [11].

In this paper, we show how a compositional mathematical systems modelling theory — which is grounded in process algebra [28, 29] and logical resource semantics [32, 34, 35, 11], which has been developed in detail by some of us elsewhere [12, 9, 10, 11], and which is supported by an execution engine and a model checker [11] — can be extended to include an account of decision-making by agents as they execute within models.

Our approach introduces into the account of processes a notion of utility that associates values to the agents' choices. Our addition of utility is formulated so as to support a key measure of decision-making in multi-agent systems: agents make their decisions in the context provided by the other agents that are executed within the model, so that different decision paths occur in different contexts. As is usual in the process-algebraic approach to modelling, the language of processes is associated with a logic of state, in the sense of Hennessy and Milner [11], in which propositional assertions describe properties of the states of the model. In this paper, these logical judgements are also context-dependent.

In Section 2, we provide a brief introduction to our background modelling theory [11] and in Section 3 we explain our utility-theoretic approach to modelling decision-making. In Section 4, we explain our contextual process calculus with utility and, in Section 5, we explain its associated logic. We conclude, in Section 6, with (a sketch of) an application of our ideas to the concept of a 'trust domain'. We provide an extended derivation of the example used in this paper in Appendix A, and full proofs of all claims in [2].

## 2. SYSTEMS MODELLING BACKGROUND

While the notion of process has been explored in some detail by the semantics community, concepts like resource have usually been treated as second class ([30] is a partial exception). From the point-of-view of a theorist, there are many advantages in doing this. We have taken the opposite view [12, 9, 10, 11]: we explore what can be gained by developing an approach in which the structures present in modelling languages are given a rigorous treatment as first-class citizens in a theory. In particular, we ensure that each component — locations, resources, and processes — is handled compositionally. These key structural components are considered, drawing upon distributed systems theory (e.g., [13]), as follows:

*Location*: Places are connected by (directed) links. Locations may be abstracted and refined provided the connectivity of the links and the placement of resources is respected. Mathematically, the axioms for locations [9] are satisfied by various graphical structures, including simple directed graphs and hyper-graphs, as well as various topological constructions [12, 9, 10, 11];

*Resource*: The notion of resource captures the components of the system that are manipulated by its processes (see below). Resources include things like the components used by a production line, the tools on a production line, computer memory, system operating staff, or system users, as well as money. Conceptually, the axioms of resources are that they can be combined and compared. Mathematically, we model this notion using *(partial commutative) resource monoids* [32, 12, 9, 10, 11]. That is, structures $\mathbf{R} = (\mathbf{R}, \sqsubseteq, \circ, e)$ with carrier set $\mathbf{R}$, preorder $\sqsubseteq$, and par-tial binary composition $\circ$ with unit $e$, and which satisfies the *bifunctoriality condition*: $R \sqsubseteq R'$ and $S \sqsubseteq S'$ and $R \circ S$ is defined implies $R' \circ S'$ is defined and $R \circ S \sqsubseteq R' \circ S'$, for all $R, S, R', S' \in \mathbf{R}$. In this paper, the order $\sqsubseteq$ is always taken to be equality. Let $\mathbf{R}$ be a given resource monoid;

---

[*]Email: g.a.anderson@abdn.ac.uk
[†]Email: matthew.collinson@abdn.ac.uk
[‡]Email: d.j.pym@abdn.ac.uk

*Process*: The notion of process captures the (operational) dynamics of the system. Processes manipulate resources in order to deliver the system's intended services. Mathematically, we use algebraic representation of processes based on the ideas in [28], integrated with the notions of resource and location [12, 9, 10, 11].

Let Act be a commutative monoid of *actions*, with multiplication written as juxtaposition and unit 1. Let $a, b \in$ Act, etc. The execution of models based on these concepts, as formulated in [12, 9, 10, 11], is described by a transition system with a basic structural operational semantics judgement [33, 28] of the form

$$L, R, E \xrightarrow{a} L', R', E',$$

which is read as 'the occurrence of the action $a$ evolves the process $E$, relative to resources $R$ at locations $L$, to become the process $E'$, which then evolves relative to resources $R'$ at locations $L'$'.

The meaning of this judgement is given by a structural operational semantics [33, 28]. The basic case, also know as 'action prefix', is the rule

$$\frac{}{L, R, a : E \xrightarrow{a} L', R', E'} \quad \mu(L, R, a) = (L', R').$$

Here $\mu$ is a 'modification' function from locations, resources, and actions (assumed to form a monoid) to locations and resources that describes the evolution of the system when an action occurs. Neglecting locations for now, partial function $\mu : \text{Act} \times \mathbf{R} \to \mathbf{R}$ is a *modification* if it satisfies the following conditions for all $a, b, R, S$: $\mu(1, R) = R$; if $R \circ S$ and $\mu(a, R) \circ \mu(b, S)$ are defined, then $\mu(ab, R \circ S) = \mu(a, R) \circ \mu(b, S)$.

There are also rules giving the semantics to combinators (which together form a a complete set for enumerating r.e. graphs) for concurrent composition, choice, and hiding — similar to restriction in SCCS and other process algebras (e.g., [28, 29] — as well for recursion. For example, the rule for synchronous concurrent composition of processes is

$$\frac{L, R, E \xrightarrow{a} L', R', E' \quad M, S, F \xrightarrow{b} M', S', F'}{L \cdot M, R \circ S, E \times F \xrightarrow{ab} L' \cdot M', R' \circ S', E' \times F'},$$

where we presume, in addition to the evident monoidal compositions of actions and resources, a composition on locations. The rules for the other combinators, with suitable coherence conditions on the modification functions, follow similar patterns [11]. Note that our choice of a synchronous calculus retains the ability to model asynchrony [28, 29, 14] (this doesn't work the other way round).

Associated with this transition semantics is a modal logic, given in the sense of Hennessy and Milner [19], with satisfaction relation $L, R, E \models \phi$ read as 'property $\phi$ holds of the process $E$ executing with resources $R$ at locations $L$'. In developing our subsequent theory, we will, for brevity and simplicity, suppress locations, working with a calculus and associated logic of resources and processes, based on judgements of the form $R, E \xrightarrow{a} R', E'$ and $R, E \models \phi$, respectively. Whilst this simplification represents some loss of generality (see [11] for more detail), the essential ideas are not significantly affected (and, indeed, some aspects of location can be coded within resource). In our discussion of the concept of a trust domain, in Section 6, the intuitive need for location, be it 'logical' or 'physical', is apparent and we revisit the concept of location there.

The logic includes — in addition to the usual additive connectives, quantifiers, and modalities that are familiar from Hennessy-Milner logic — multiplicative connectives, quantifiers, and modalities [12, 9, 10, 11]. For example, dropping locations for brevity, multiplicative conjunction is defined by the logical decomposition of the system, as follows:

$$R, E \models \phi_1 * \phi_2 \quad \text{iff} \quad \begin{array}{l} \text{there are } R_1, R_2 \text{ and } E_1, E_2 \text{ s.t.} \\ R = R_1 \circ R_2, E \sim E_1 \times E_2, \text{ and} \\ R_1, E_1 \models \phi_1 \text{ and } R_2, E_2 \models \phi_2. \end{array}$$

Here $\sim$ is bisimulation, explained in detail for this set-up in [12, 9, 10, 11], where treatments of location can also be found.

The action modalities work just as in Hennessy–Milner logic. For example,

$$R, E \models \langle a \rangle \phi \quad \text{iff} \quad \begin{array}{l} \text{for some } R', E' \text{ s.t. } R, E \xrightarrow{a} R', E' \\ \text{and } R', E' \models \phi. \end{array}$$

The multiplicative version of this rule would permit the evolution by action $a$ to employ additional resource; that is, for some $S$, $R'$, and $E$, $R \circ S, E \xrightarrow{a} R', E'$. Details and theoretical development may be found in [12, 9, 10, 11]. Although the basic logic of bunched implications, with intuitionistic additives and multiplicatives, is decidable [17], its counterpart with classical additives, widely known as 'Boolean BI' and the basis of Separation Logic [37], is not [6, 18]. We conjecture that the undecidability of Boolean BI implies the undecidability of the logic presented here.

In addition to the structural components of models, we consider also the environment within which a system exists:

*Environment*: All systems exist within an external environment, which is typically treated as a source of events that are incident upon the system rather than being explicitly stated. Mathematically, environments are represented stochastically, using probability distributions that are sampled in order to provide such events [12, 9, 10, 11].

The modal logic discussed above can also be extended to the stochastic world, but an account of this is beyond our present needs and scope. Related work can be found in [38]. Logical reasoning about distributed systems has also been studied by Barwise and Seligman [4]. We provide full proofs of all claims made in this paper in the accompanying technical report [2].

## 3. MODELLING DECISION-MAKING

One use of the process component within our models is to represent agents within a system as they explore their worlds, making decisions between the choices that are available to them. In the set-up described so far, as presented in [12, 9, 10, 11], we have considered (again, suppressing location, for brevity) an operational rule for choice of the form

$$\frac{R, E_i \xrightarrow{a} R', E'}{R, \sum_{i \in I} E_i \xrightarrow{a} R', E'},$$

where $I$ is an indexing set. This rule is understood, in the style of structural operational semantics [33], read from con-clusion to premisses, as follows: the sum of the processes can evolve by an action $a$, to become $R', E'$ if one of its summands can evolve by the action $a$ to become $R', E'$. In other words, there is a set of possible evolutions each element of which leads to the evolution of the sum.

Models of distributed systems often capture situations in which an agent or group of agents is exploring a world and interacting with it and themselves. In such cases, a process containing choices of the kind discussed represents the choices made by agents as they evolve. It is therefore useful to extend our modelling theory to provide an account of agents' decision-making.

Our approach, applying our methodology of incorporating representations of system modelling components as first-class citizens, is to model agents' decision-making by developing a utility-carrying version of the location-resource-process calculus sketched above. It is possible to encode some notions of location in the resource component; for brevity and simplicity, we follow this approach to location for the remainder of this paper. We return, in particular, to the strengths of using location in modelling when we consider trust domains in Section 6.

The key idea is to replace the simple choice combinator described above with the *utility-dependent choice* (or simply *sum*) $\sum_{i \in I} {}_u E_i$, in which an agent has a choice between alternatives $E_i$, and its preference is codified by the utility $u \in U$. The operational rules of the calculus ensure that this preference takes into account the wider context in which the choice is embedded. For example, if the choice $R, a:E +_u b:F$ (here we use an infix notation) occurs within a wider context $R \circ S, (a:E +_u b:F) \times G$, then preference will be determined by the utility calculations

$$u(R \circ S, a:E \times G) \quad \text{and} \quad u(R \circ S, b:F \times G). \qquad (1)$$

If the former is greater than the latter, the $a:E$ option will be chosen. An occurrence of the same choice $R, a:E +_u b:F$ within a different context, such as $R \circ T, (a:E +_u b:F) \times H$ with $G \neq H$ or $S \neq T$, will have different utility calculations, and may result in $b:F$ being chosen instead.

Many process calculi include a form of prioritized sum, for example [41]. In prioritized sums, say $w \cdot a:E + w' \cdot b:F$, with $w > w'$, the option $a:E$ is always chosen in any context in which both $a$ and $b$ are available (which they may not be, because of restriction operations). In contrast, our utility-based sum can make different choices in different contexts even when the same options are available.

In our set-up, resource-process contexts correspond to the semantical notion of world. Utility functions are simply given, being aspects of agents described in the model by the modeller. A *preference order* $\succeq_u$ on worlds is induced, in the usual way, by: $R, E \succeq_u S, F$ iff $u(R, E) \geq u(S, F)$, for all $R, E$ and $S, F$. This means that the mathematical structures we are considering appear to be similar to those used to give a semantics for dynamic logics of preference [42], but in a special case where processes are used to give a very richly descriptive dynamics.

The use of a utility restricts the ordinal preference relations that can be represented (in the usual way [25]), but it is a trivial step to replace the utility comparisons in our calculus with more general relational comparisons, if so required.

Our utility calculations are not required to be determined by the system dynamics described by the transition system. At a sum, an agent makes a decision as to its own (partial) control (i.e., process) choice knowing its context, but not using information about the resulting future evolution of the system. Even if an agent may face a sequence of decisions, we are not forced to model it as undertaking a traditional, rational, multi-stage decision process. The decoupling of choice from dynamics should also make it possible to incorporate expected utility for probabilistic choices [41] in a natural fashion. Traditional decision theory usually treats decision situations in a flat, atomic way, and is not concerned with similar choice-points in different contexts. Prior works that address this issue include [21, 16].

Process calculi in which contexts are treated as first-class citizens include [8, 40, 7, 39]. Logics of propositions in context have also been extensively studied, for example [27]. Points of contact between decision theory and process calculus include [15, 31, 3]. Our approach differs from these in having an explicit utility-based choice constructor, which, in particular, which takes into account the wider context. The importance of the combination of utility and process in reasoning about trust has also been recognized in [3].

## 4. A PROCESS ALGEBRA WITH UTILITY

In process calculi such as the one sketched in Section 1, the behaviour of a composite process, such as $E_1 \times E_2$, is usually defined in terms of the behaviours of its sub-processes, with the reductions of $E_1$ and $E_2$ being independent of each other. In our calculus, however, choices take account of the context in which sub-processes are reduced, so that the reductions of $E_1$ and $E_2$ may not be independent.

To see how this works, consider the example used in the utility expressions (1) in Section 3. The process $a:A+_u b:B$ reduces taking account of its context $G$. We annotate the context in which a process is reduced on the underside of the reduction arrow (e.g., $R, a:A +_u b:B \xrightarrow[S,[\,]\times G]{a} R', A$), where $[\,]$ denotes the hole into which $a:A +_u b:B$ may be substituted to regain the complete system $(a:A +_u b:B) \times G$, and $S$ are the resources allocated to $G$. Note also that any choices in $[\,] \times G$ may depend on what process is substituted into the hole $[\,]$. We therefore annotate the process that is substituted, into the process being reduced, on top of the reduction arrow; for example, $S, [\,] \times G \xrightarrow{R, a:A+_u b:B}^b S', [\,] \times G'$.

So, the key judgement of the reduction relation for processes with utility is of the form

$$C \xrightarrow[C_1]{C_2}^a C', \qquad (2)$$

which denotes how a context $C$, that exists in a system that can be decomposed as $C_1(C(C_2))$, reduces. We refer to $C$ as the *(primary) context*, $C_1$ as the *outer context*, and $C_2$ as the *substituted*, or *inner context*. Intuitively this denotes the reduction of one part, $C$, of an entire system, $C_1(C(C_2))$. In order to reason compositionally we wish to be able to describe the reduction of C independently and structurally. As choices can take account of context, this is not possible. The choices in $C$, however, only make use of the definition of $C_1$ and $C_2$, and disregard their structure. Hence we do not need to reason over the structure of $C_1$ and $C_2$, as we do with $C$, only to record their definitions which can then be referred to at choice points. They are therefore annotated on the reduction arrow for reference, but are not reduced in said relation.

We now describe the theoretical set-up in detail. Assume a set $U$ of symbols, called *formal utilities*, with a distinguished element $0_U$, called the *neutral utility*. *Processes* are generated by the grammar:

$$E ::= \mathbf{1} \mid [\,] \mid a : E \mid \sum_{i \in I}{}_u E_i \mid E \times E.$$

These are really process contexts: the term $[\,]$ is a *hole* into which other processes may be substituted. For this work, it turns out to be convenient to develop contexts as first-class citizens rather than merely meta-theoretic tools.

The *choice* $\sum_{i \in I}{}_u E_i$ is new: it describes situations in which an agent has a choice between alternatives $E_i$ indexed by a $i \in I$, and its preference (in a larger context) is codified by the utility $u \in U$. The infix operator $E +_u F$ may be used for binary sums, and the subscript $u$ may be dropped when $u = 0_U$. The *zero* process $\mathbf{0}$ is defined to be the sum indexed by the empty set and the neutral utility. The zero process, *unit* process $\mathbf{1}$, and *synchronous products* $E \times F$ are well-known in process calculus, as are *prefixes* $a : E$, where $a \in \text{Act}$.

A process $E$ is *well-formed* if it contains at most one hole and that hole is not guarded by action prefixes. The process $E$ is *closed* if it has no holes and *open* otherwise. Let $PCont$ be the set of all well-formed processes, $PCCont$ be the set of all closed well formed processes, and $POCont$ be the set of all open well-formed processes.

Let $\mathbf{R}$ be a resource monoid and $\mu$ be a fixed modification function, as defined in Section 2. Define the products of sets $Cont = \mathbf{R} \times PCont$, $CCont = \mathbf{R} \times PCCont$ and $OCont = \mathbf{R} \times POCont$. The letter $C$ is reserved for contexts. Define $C_\emptyset = e, [\,]$. Brackets will be freely used to disambiguate both processes and contexts. For $C = R, E$, the notational abuses $C \times F = R, (E \times F)$ and $C +_u F = R, (E +_u F)$ will sometimes be used. Substitution in processes, $E(F)$, replaces all occurrences of $[\,]$ in $E$ with $F$; for example, $(([\,] +_u E) \times G)(F) = (F +_u E) \times G$. Substitution of contexts $C_1(C_2)$, where $C_1 = R, E$ and $C_2 = S, F$, is defined as follows: if $E$ is open, then $C_1(C_2) = R \circ S, E(F)$, where $E(F)$ is process substitution; if $E$ is closed, then $C_1(C_2) = C_1$.

We assume that, for each formal utility $u \in U$, there is an associated, real-valued *utility function* $u : Cont \longrightarrow \mathbb{R}$ [24] that fixes an interpretation for each formal symbol $u \in U$. The identically zero function is associated with $0_U$. Henceforth, we do not distinguish between formal utilities and their utility functions.

The operational semantics of our process-utility calculus is given in Figure 1. The side-condition $(S\sum)$ is that $C_3 = C_1((e, \sum_{I \setminus j}{}_u E_i(C_2)) +_u [\,])$ and $\forall i \in I.u(C_1(R, E_i(C_2))) \leq u(C_1(R, E_j(C_2)))$. The side-condition $(S\times)$ is that $C_3 = C_1((S, F(C_2)) \times [\,])$ and $C_4 = C_1((R, E(C_2)) \times [\,])$.

The unit process always ticks, effecting no change. The prefix process evolves via its head action. The hole rule is a technical one used to terminate reduction derivations of open contexts. The sum process $\sum_{I}{}_u E_i$ represents a preference-based choice by the agent: it follows the behaviour of any of its constituent $E_j$ which has at least as high a value ascribed by its utility $u$ as any other option $E_i$ for $i \in I$. The

$$\overline{R, \mathbf{1} \xrightarrow[C_1]{C_2}{}^1 R, \mathbf{1}}\quad (\textsc{Tick})$$

$$\overline{R, a : E \xrightarrow[C_1]{C_2}{}^a \mu(a, R), E}\quad (\textsc{Prefix})$$

$$\frac{C_2 \xrightarrow[C_1]{(e, \mathbf{1})^a} C_2'}{e, [\,] \xrightarrow[C_1]{C_2}{}^1 e, [\,]}\quad (\textsc{Hole})$$

$$(S\sum)\quad \frac{R, E_j \xrightarrow[C_3]{C_2}{}^a S, F}{R, \sum_{I}{}_u E_i \xrightarrow[C_1]{C_2}{}^a S, F}\quad (\textsc{Sum})$$

$$(S\times)\quad \frac{R, E \xrightarrow[C_3]{C_2}{}^a R', E' \quad S, F \xrightarrow[C_4]{C_2}{}^b S', F'}{R \circ S, E \times F \xrightarrow[C_1]{C_2}{}^{ab} R' \circ S', E' \times F'}\quad (\textsc{Prod})$$

Figure 1: Operational Semantics

first special case of the sum is for the zero process $\mathbf{0}$, which never evolves. The second special case is where $u = 0_U$ and the sum becomes an ordinary non-deterministic sum: in this case, the utility is irrelevant, and the sum may evolve as any of its component processes. The product evolves two processes synchronously in parallel, according to the decomposition of the associated resources. An important feature of this system is that contextual information about conclusions is propagated up to premises. In the product case, information about each premiss is propagated up from the conclusion to the other premiss, so that derivations of transitions occur in context.

To demonstrate how contextual decisions can be utilized in modelling, we give a simple example (inspired by [5]). Consider a banker who has a presentation (for a client, that includes confidential business data) on a USB drive. The banker may chose to access the drive or not, depending on the situation. The banker is modelled as a process

$$Banker = present : Banker' +_{u_B} idle_B : Banker', \quad (3)$$

where $u_B$ represents its preferences. The banker may be willing to access the presentation when visiting a client, on the assumption that the client's network is firewalled, so making the document safe from attack. In order to do so, however, the banker must be given access to a computer by the client. The client is modelled as

$$Client = logIn : Client' +_{0_U} idle_C : Client', \quad (4)$$

which, for simplicity, makes a non-deterministic choice between logging the guest in and idling. The interaction between the banker and the client is a form of joint access control, in which the banker cannot show the presentation without having been logged in, and the client cannot see the presentation unless the banker accesses it. Here we can show that the principals co-operate to access the presentation:

$$R, Client \times Banker \xrightarrow{logIn,present} S, Client' \times Banker'. \tag{5}$$

If the banker's utility is $u_B$, then we have

$$u_B(R, Client \times (idle_B : Banker')) \leq$$
$$u_B(R, Client \times (present : Banker')). \tag{6}$$

In a different situation — here, a different context — the banker may make a different decision. The banker may use a home computer, compromised by an attacker, who wants to steal the presentation, but cannot unless the banker accesses it from the USB stick. The attacker is modelled as

$$Attacker = steal : Attacker' +_{0_U} idle_A : Attacker'. \tag{7}$$

In this situation, the banker prefers to idle than to work on the presentation. As, in order for the attacker to steal the presentation the banker must access it, and the banker choses not to, then the attacker must also idle, so that

$$R, Attacker \times Banker \xrightarrow{idle_A, idle_B} S, Attacker' \times Banker'. \tag{8}$$

Here the banker's utility yields

$$u_B(R, Attacker \times (present : Banker')) \leq$$
$$u_B(R, Attacker \times (idle_B : Banker')). \tag{9}$$

A more detailed exposition of these examples is presented in Appendix A.

A fundamental aspect of process calculus is the ability to reason equationally about behavioural equivalence of processes [28]. We now adapt these notions to suit the calculus above, which incorporates ideas from [12, 9, 11].

The *bisimilarity (or bisimulation) relation* $\sim \; \subseteq PCont \times PCont$ is the largest binary relation such that, if $E \sim F$, then $\forall a \in$ Act, $\forall R, R', S, T \in \mathbf{R}$, and for all $G, H, I, J \in PCont$ with $G \sim I$ and $H \sim J$, then

1. $\forall E' \in PCont$, if $R, E \xrightarrow[S,G]{T,H}{}^a R', E'$ then $\exists F'$ such that $R, F \xrightarrow[S,I]{T,J}{}^a R', F'$ and $E' \sim F'$, and

2. $\forall F' \in PCont$, if $R, F \xrightarrow[S,I]{T,J}{}^a R', F'$ then $\exists E'$ such that $R, E \xrightarrow[S,G]{T,H}{}^a R', E'$ and $E' \sim F'$.

The union of any set of relations that satisfy these two conditions also satisfies these conditions, so the largest such relation is well-defined. Define $\sim \; \subseteq Cont \times Cont$ by: if $E \sim F$ then $R, E \sim R, F$ for all $R \in \mathbf{R}$ and $E, F \in Cont$.

DEFINITION 1. *A utility, $u$, respects bisimilarity if, for all $C_1, C_2 \in Cont$, $C_1 \sim C_2$ implies $u(C_1) = u(C_2)$.*

That is, behaviourally equivalent (bisimilar) states are required to be indistinguishable by $u$. The set $U$ of utilities respects bisimilarity if every $u \in U$ respects bisimilarity. Henceforth utilities are assumed to respect bisimilarity. We can show that if bisimilar contexts are substituted into each other, then the result is bisimilar:

PROPOSITION 1. *If $E \sim G$ and $F \sim H$, then $E(F) \sim G(H)$.*

We can then prove a key property for reasoning compositionally.

THEOREM 1 (BISIMULATION CONGRUENCE). *The relation $\sim$ is a congruence. It is reflexive, symmetric and transitive, and for all $a, E, F, G$ with $E \sim F$, and all families $(E_i)_{i \in I}$, $(F_{i \in I})_I$ with $E_i \sim F_i$ for all $i \in I$, $a : E \sim a : F$, $E \times G \sim F \times G$, and $\sum_{i \in I}{}_u E_i \sim \sum_{i \in I}{}_u F_i$.*

PROOF. Symmetry, reflexivity, and transitivity are straightforward. Prefixed processes $a : E$ and $a : F$ can only reduce via an $a$ action to $E$ and $F$, which are bisimilar.

Consider the choices $E +_u G$ and $F +_u G$, in (bisimilar) outer contexts $C_1$ and $C_2$, with (bisimilar) inner contexts $C_3$ and $C_4$, and the case where $R, E +_u G \xrightarrow[C_1]{C_2}{}^a S, E'$. By the (SUM) rule we know that $u(C_1(R, G(C_2))) \leq u(C_1(R, E(C_2)))$ and that $R, E \xrightarrow[C_5]{C_2}{}^a S, E'$, where $C_5 = C_1((e, G(C_2)) +_u [\,])$. Let $C_6 = C_3((e, G)(C_4) +_u [\,])$; we can show that $C_5 \sim C_6$ (by Proposition 1), and hence that $R, F \xrightarrow[C_6]{C_4}{}^a S, F'$. By Proposition 1 we know that $C_1(R, E(C_2)) \sim C_3(R, F(C_4))$. Using Proposition 1, and the fact that utility functions respect bisimilarity we can show that $u(C_3(R, G(C_4))) \leq u(C_3(R, F(C_4)))$, and hence that $R, F +_u G \xrightarrow[C_3]{C_4}{}^a S, F'$.

The product case follows from the fact that the contexts in which each sub-process reduces, such as $C_1((S, G(C_2)) \times [\,])$ for $E$, is bisimilar to the context in which the counterpart reduces, for example $C_3((S, G(C_4)) \times [\,])$ for $F$, by Proposition 1. □

In order to reason equationally about processes, it is also useful to establish various algebraic properties concerning parallel composition and choice. We derive these below, for our calculus. In order to do so, we make some additional definitions concerning utility functions.

DEFINITION 2. *The set of utilities, $U$, is (algebraically) accordant it respects bisimilarity and, for all $u, v \in U$, all $C, C_1, C_2, C_3, C_4 \in Cont$, all $E, F, G \in PCont$, and $R \in \mathbf{R}$,*

1. *$u(C(R, F)) \leq u(C(R, E))$ and $u(C(R, G)) \leq u(C(R, E))$ if and only if $u(C(R, F +_v G)) \leq u(C(R, E))$,*

2. *$u(C(R, F)) \leq u(C(R, E))$ and $u(C(R, G)) \leq u(C(R, E))$ if and only if $u(C(R, G)) \leq u(C(R, E +_u F))$.*

3. *for all $R, E$, $u(C(R, 0)) \leq u(C(R, E))$, and*

4. *for all $C_1 \sim C_3, C_2 \sim C_4, R, E, F, G$, $u(C_1(R, E \times G +_u F \times G(C_2))) = u(C_3(R, (E +_u F) \times G(C_4)))$.*

We use the binary version of sum here in order to aid comprehension, but finite choices between sets of processes work straightforwardly. Any real-valued function defined on the quotient $Cont/\sim$ defines a utility that respects bisimilarity.

PROPOSITION 2 (ALGEBRAIC PROPERTIES). *If $U$ is accordant, then:*

| | | | |
|---|---|---|---|
| 1 | $E +_u F$ | $\sim$ | $F +_u E$ |
| 2 | $E +_u (F +_u G)$ | $\sim$ | $(E +_u F) +_u G$ |
| 3 | $E +_u \mathbf{0}$ | $\sim$ | $E$ |
| 4 | $E \times \mathbf{0}$ | $\sim$ | $\mathbf{0}$ |
| 5 | $E \times \mathbf{1}$ | $\sim$ | $E$ |
| 6 | $E \times F$ | $\sim$ | $F \times E$ |
| 7 | $E \times (F \times G)$ | $\sim$ | $(E \times F) \times G$ |
| 8 | $(E +_u F) \times G$ | $\sim$ | $E \times G +_u F \times G$. |

PROOF. The interesting cases are: associativity of choice (2) as it uses Definition 2.1 and 2.2, the unit of choice (3) as it uses Definition 2.3, and distributivity of product over choice (8), which uses Definition 2.4. The others are as usual, and do not make use of the accordance properties.

As a representative of the interesting properties we prove the associativity of choice. Consider the choices $E +_u (F +_u G)$ and $(E +_u F) +_u G$, in (bisimilar) outer contexts $C_1$ and $C_2$, with (bisimilar) inner contexts $C_3$ and $C_4$, and the case where $R, E +_u (F +_u G) \xrightarrow[C_1]{C_2}{}^a S, E'$. By the (SUM) rule we know that $u(C_1(R, F +_u G(C_2))) \leq u(C_1(R, E(C_2)))$. By the accordance properties (Def. 2.1) we then know that $u(C_1(R, F(C_2))) \leq u(C_1(R, E(C_2)))$ and $u(C_1(R, G(C_2))) \leq u(C_1(R, E(C_2)))$.

Let $C_5 = C_1((e, F +_u G(C_2)) +_u [\,])$ and $C_6 = C_3((e, F +_u G(C_4)) +_u [\,])$; using Proposition 1 we can show that these two contexts are bisimilar. By the (SUM) rule we know that, as $R, E +_u (F +_u G) \xrightarrow[C_1]{C_2}{}^a S, E'$, then $R, E \xrightarrow[C_5]{C_2}{}^a S, E'$. So, by the definition of bisimulation, we have that $R, E \xrightarrow[C_6]{C_4}{}^a S, E'$.

As utility respects bisimilarity (Definition 1) we have that $u(C_3(R, G(C_4))) = u(C_1(R, G(C_2))) \leq u(C_1(R, E+_uF(C_2))) = u(C_3(R, E+_u F(C_4)))$, and that $u(C_3(R, F(C_4))) = u(C_1(R, F(C_2))) \leq u(C_1(R, E(C_2))) = u(C_3(R, E(C_4)))$. Let $C_7 = C_3((e, G(C_4)) +_u [\,])$. By (SUM) we can then show that $R, E +_u F \xrightarrow[C_7]{C_4}{}^a S, E'$, and finally that $R, (E +_u F) +_u G \xrightarrow[C_3]{C_4}{}^a S, E'$. □

Future work includes extending the calculus to include probabilistic choice [41] and expected utility [24]. It would also be interesting to consider whether the (pre)sheaf-theoretic semantics considered by Winskel [23] can be adapted to our calculus.

## 5. A PROCESS LOGIC WITH UTILITY

We now introduce a modal logic of system properties. The semantics is given using a satisfaction relation

$$C \vDash_{C'} \phi,$$

where $C$ is a closed context, $C'$ is a context and $\phi$ is a formula of a (Hennessy–Milner-style) modal logic of processes: this may be read 'the *primary context* $C$ satisfies $\phi$ in the *surrounding context* $C'$' (cf. (2)). The context $C$ may satisfy different logical propositions, perhaps even negations of each other, when placed in different surrounding contexts; an example of this is below. Context-sensitive logics have been studied by other authors [26, 4]. The structural nature of processes and resources provides a semantic framework in which such logics seem particularly natural.

The propositions of the logic are defined by the grammar

$$\phi ::= \ p \mid \bot \mid \top \mid \neg\phi \mid \phi \wedge \phi \mid \phi \vee \phi \mid \phi \to \phi \mid \\ \langle a \rangle \phi \mid [a]\phi \mid I \mid \phi * \phi \mid \phi \mathbin{-\!*} \phi,$$

where $p$ ranges over atomic propositions, and $a$ over actions. The symbols for propositions for *truth, falsehood, negation* and *(additive) conjunction, disjunction,* and *implication* are standard. The *(additive) modal connectives* are $\langle a \rangle$ and $[a]$. The connectives $I$, $*$, and $-\!*$ are the *multiplicative unit, conjunction,* and *implication*, respectively.

A *valuation*, $\mathcal{V}$, is a function that maps each atomic proposition to a $\sim$-closed set of closed contexts. The satisfaction relation is specified in Figure 2.

In the interpretation of atoms, the surrounding context is wrapped around the primary context, and the valuation of the atom consulted to see if it contains this compound context. This is what makes our logic context-sensitive. $\top$, $\bot$, $\neg$, $\wedge$, $\vee$, and $\to$ are all interpreted (essentially) classically.

The interpretation of the multiplicatives connectives here is similar to that for the logic MBI in [12]. Recall also the comments on $*$ in Section 1. The semantics of $*$ in Figure 2 is slightly modified, because of the way that contextual information is propagated upwards from conclusion to premisses in the product rule of the operational semantics.

The standard interpretation of Hennessy–Milner logics uses the relation specified by the operational semantics as a Kripke structure to support the modal connectives. In our work, the operational semantics is more complex: a context occurs, and reduces alongside an outer context. Hence when we consider whether $C_1 \vDash_{C_2} \langle a \rangle \phi$ holds, we have to consider whether there are reductions of the form $C_1 \xrightarrow[C_2]{C_\emptyset}{}^a C_1'$ and $C_2 \xrightarrow[C_\emptyset]{C_1}{}^b C_2'$ such that $C_1' \vDash_{C_2'} \phi$. The occurrences of the empty context ensure that no extraneous contextual information is introduced into the reductions of interest. The $[a]$ modality is interpreted similarly.

Recall the example of the banker who decides which actions to take in different contexts (3-9). In a situation that consists of a client (context $C_C$), the banker chooses to access the presentation, but in a situation that consists of an attacker (context $C_A$) the banker chooses not to: that is,

$$\begin{array}{c} R_B, Banker \vDash_{C_C} \langle present \rangle \top \\ R_B, Banker \vDash_{C_A} \neg \langle present \rangle \top. \end{array} \quad (10)$$

A derivation of these properties is in Appendix A. Hence, in different contexts the process satisfies different propositions that, moreover, would be inconsistent over the same context.

Behaviourally equivalent processes are also logically equivalent (they satisfy the same logical properties). This is half of the Hennessy–Milner property [19, 20].

THEOREM 2. *If $C_1 \vDash_{C_2} \phi$, $C_1 \sim C_3$, and $C_2 \sim C_4$, then $C_3 \vDash_{C_4} \phi$.*

PROOF. A standard argument, by induction over the definition of $C_1 \vDash_{C_2} \phi$, using Proposition 1 in the cases where contexts are extended. □

Hence, bisimilar processes can be used interchangeably within a larger system, without changing the logical properties of the larger system.

It is unclear whether a useful converse can be obtained. With restrictions on the available fragments of the logic, and a different equivalence relation, however, it is possible to obtain a converse. To this end, we introduce the *local equivalence relation* $\approx \ \subseteq (OCont \times Cont \times CCont) \times (OCont \times Cont \times CCont)$, the largest binary relation such that, if $C_1, A, D_1 \approx C_2, B, D_2$, $\forall a \in$ Act where $A = R, E$ and $B = S, F$ (with $A'$ and $B'$, etc., modifying them, as usual), then

1. $\forall C_1' \in OCont, A' \in Cont, D_1' \in CCont$, if $A \xrightarrow[C_1]{D_1}{}^a A'$ and $C_1 \xrightarrow[C_\emptyset]{A(D_1)}{}^c C_1'$ and $D_1 \xrightarrow[C_1(A)]{C_\emptyset}{}^d D_1'$ then $\exists C_2' \in$

| | | | | | | |
|---|---|---|---|---|---|---|
| $C \vDash_{C'} p$ | iff | $C'(C) \in \mathcal{V}(p)$ | | $C_1 \vDash_{C_2} \langle a \rangle \phi$ | iff | $\exists C_1', C_2', b$ such that if $C_1 \xrightarrow[C_2]{C_\emptyset}{}^a C_1'$ and $C_2 \xrightarrow[C_\emptyset]{C_1}{}^b C_2'$, then $C_1' \vDash_{C_2'} \phi$ |
| $C \vDash_{C'} \bot$ | | never | | | | |
| $C \vDash_{C'} \top$ | | always | | $C_1 \vDash_{C_2} [a] \phi$ | iff | $\forall C_1', C_2', b$ such that if $C_1 \xrightarrow[C_2]{C_\emptyset}{}^a C_1'$ and $C_2 \xrightarrow[C_\emptyset]{C_1}{}^b C_2'$, then $C_1' \vDash_{C_2'} \phi$ |
| $C \vDash_{C'} \neg \phi$ | iff | $C \nvDash_{C'} \phi$ | | | | |
| $C \vDash_{C'} \phi \wedge \psi$ | iff | $C \vDash_{C'} \phi$ and $C \vDash_{C'} \psi$ | | $R, E \vDash_{C'} I$ | iff | $R = e$ and $E \sim \mathbf{1}$ |
| $C \vDash_{C'} \phi \vee \psi$ | iff | $C \vDash_{C'} \phi$ or $C \vDash_{C'} \psi$ | | $R, E \vDash_{C'} \phi * \psi$ | iff | $\exists S, T, F, G$ such that $R = S \circ T$, $E \sim F \times G$, and $S, F \vDash_{C'(T, [] \times G)} \phi$ and $T, G \vDash_{C'(S, F \times [])} \psi$ |
| $C \vDash_{C'} \phi \to \psi$ | iff | $C \vDash_{C'} \phi$ implies $C \vDash_{C'} \psi$ | | $R, E \vDash_{C'} \phi \twoheadrightarrow \psi$ | iff | $\forall S, F$ such that $R \circ S$ is defined and $S, F \vDash_{C'} \phi$, $R \circ S, E \times F \vDash_{C'} \psi$ |

Figure 2: Interpretation of Propositional Formulae

$OCont$, $B' \in Cont$, $D_2' \in CCont$ such that $B \xrightarrow[C_2]{D_2}{}^a B'$ and $C_2 \xrightarrow[C_\emptyset]{B(D_2)}{}^c C_2'$ and $D_2 \xrightarrow[C_2(B)]{C_\emptyset}{}^d D_2'$ and $C_1', A', D_1' \approx C_2', B', D_2'$

2. $\forall C_2' \in OCont$, $B' \in Cont$, $D_2' \in CCont$, if $B \xrightarrow[C_2]{D_2}{}^a B'$ and $C_2 \xrightarrow[C_\emptyset]{B(D_2)}{}^c C_2'$ and $D_2 \xrightarrow[C_2(B)]{C_\emptyset}{}^d D_2'$ then $\exists C_1' \in OCont$, $A' \in Cont$, $D_1' \in CCont$ such that $A \xrightarrow[C_1]{D_1}{}^a A'$ and $C_1 \xrightarrow[C_\emptyset]{A(D_1)}{}^c C_1'$ and $D_1 \xrightarrow[C_1(A)]{C_\emptyset}{}^d D_1'$ and $C_1', A', D_1' \approx C_2', B', D_2'$

3. $R = S$.

The union of any set of relations that satisfy these two conditions also satisfies these conditions, so the largest such relation is well-defined. We define $C_1, A \approx C_2, B$ whenever $C_1, A, D \approx C_2, B, D$, for all $D$.

Fundamentally, this equivalence relation starts from the view that processes should be considered equivalent whenever they have the same behaviour given the same resources and context. The local equivalence relation fails to be a congruence, however, as it is not respected by the product constructor, $\times$, for processes [9]. Therefore, we do not have an analogue of Theorem 1 for local equivalence. (Note that, in [12, 9, 11], the equivalence corresponding to the equivalence $\sim$ taken here is referred to as the *global equivalence*.)

We can, however, obtain a version of the full Hennessy-Milner theorem, provided we restrict the logic to the fragment without $\twoheadrightarrow$. The need for this restriction arises from the failure of the local equivalence to be a congruence, because the satisfaction relation for $\twoheadrightarrow$ requires that two subsystems be combined using $\times$.

Consider the fragment of the logic that excludes $\twoheadrightarrow$. Assume that all atomic propositions are values as sets of contexts that are also closed under $\approx$. Alter the $I$ and $*$ clauses of the interpretation so that

| | | |
|---|---|---|
| $C \vDash_{C_1} I$ | iff | $C_1, C \approx C_1, (e, \mathbf{1})$ |
| $C \vDash_{C_1} \phi * \psi$ | iff | $\exists S, T$ and $F, G$ such that $C_1, C \approx C_1, (S \circ T, F \times G)$, and $S, F \vDash_{C_2} \phi$ and $T, G \vDash_{C_3} \psi$, where $C_2 = C'(T, [] \times G)$ and $C_3 = C'(S, F \times [])$. |

Define two contexts (with accompanying outer contexts) to be logically equivalent if they satisfy exactly the same set of logical statements; that is, $C_1, A \equiv C_2, B$ if and only if, for all $\phi$, $A \vDash_{C_1} \phi$ iff $B \vDash_{C_2} \phi$. The following version of Theorem 2 then holds:

THEOREM 3. *If $C_1, A, D_1 \approx C_2, B, D_2$, then $C_1, A(D_1) \equiv C_2, B(D_2)$.*

PROOF. Standard, by induction over the definition of $A(D_1) \vDash_{C_1} \phi$, using a forward-only analogous version of Proposition 1 for $\approx$, in the cases where contexts are extended. □

We can now also obtain a converse, for the local equivalence relation. Define a context to be *image finite* if it has finitely many immediate derivatives (for any given inner and outer contexts with which it reduces). We then have the following.

THEOREM 4. *If $C_1, A \equiv C_2, B$, then there exist $A_1$, $D_1$, $B_1$, $D_2$ such that $A = A_1(D_1)$, $B = B_1(D_2)$, and $C_1, A_1, D_1 \approx C_2, B_1, D_2$.*

PROOF. By contradiction. Take the finite set of contexts $\mathcal{C}$ that can be obtained through the reduction of $B_1(D_2)$ in outer context $C_2$ (with an empty inner context). If this set is empty, then we can show that $A_1(D_1) \vDash_{C_1} \langle a \rangle \top$ and $B_1(D_2) \nvDash_{C_2} \langle a \rangle \top$, which contradicts the premiss that $C_1, A \equiv C_2, B$. If the set is non-empty, then we can construct characteristic formulae $\phi_i$ for each context in $\mathcal{C}$, such that the result of reducing $A_1(D_1)$ in $C_1$ satisfies the formula, but the result of reducing $B_1(D_2)$ in $C_2$ does not. We can combine these to show that $A_1(D_1) \vDash_{C_1} \langle a \rangle (\phi_1 \wedge ... \wedge \phi_n)$ and $B_1(D_2) \nvDash_{C_2} \langle a \rangle (\phi_1 \wedge ... \wedge \phi_n)$, which again contradicts the premiss that $C_1, A \equiv C_2, B$. □

We remark that the usefulness of this result is limited by the failure of local equivalence to be a congruence. It is a strictly local reasoning tool.

Since each utility function $u$ induces a preference relation $\preceq_u$ on closed contexts, the language could easily be enriched with preference modalities such as $\langle \preceq_u \rangle$ and $[\preceq_u]$ in the style of dynamic preference logic [42]. Consider the necessitated formula, $[\preceq_u]\phi$, which denotes that any context that is valued at least as much as the current context (in the outer context) satisfies property $\phi$. Formally, this is interpreted as

$C \vDash_{C'} [\preceq_u]\phi$  iff  for all $C''$, $C'(C) \preceq_u C'(C'')$ implies $C'' \vDash_{C'} \phi$.

These modalities can interact powerfully with existing structural operators.

In game-theoretic approaches to security, the notion of a level of security is important. That is, if a defender chooses to perform some defensive action, then no matter what a given attacker does, the defender is guaranteed to maintain at least a certain level of security. With preference modalities we can make statements relevant to security levels. For example, if a defensive measure $d$ is in place, then every better state for an attacker (which would be chosen by the attacker) involves not attacking. To see this, consider the proposition

$$\phi \mathbin{-\!\ast} [d][\preceq_v](\neg\langle a \rangle \top),$$

where the attacker is characterized by the property $\phi$ and has preference function $v$. The multiplicative implication operator permits us to reason about substitution within arbitrary contexts, and hence of the efficacy of defensive measures relative to an arbitrary (partially) described attacker.

The logic might also be enriched to handle expected utility [24]. Quantitative path-based logical properties of Markov Chains are studied in [22]: they can reason about complex notions, such as average utility with a given time discount, but do not provide compositionality results over model structures. A more extensive study of such extensions is future work.

## 6. TRUST DOMAINS

An agent, situated within a system that contains also other agents, may establish a part of the system, or a collection of other agents within the system, that it trusts. Similarly, a system's designer or manager might establish a collection of parts of the system such that, within any given part, the agents trust one another. We shall refer to such a part of the system, or such a collection of agents, as a 'trust domain'.

The term 'trust domain' is in use in range of settings, such as the Trusted Computing Project (www.trustedcomputing.org.uk), the Open Trusted Computing (OpenTC) consortium (www.opentc.net), and the 'Trust Domains' project (www.hpl.hp.com/research/cloud_security/TrustDomains.pdf). The literature on models of trust is very large and cannot be surveyed in this short article, but a good survey with a relevant perspective for us is [36].

In this section, we consider how the process-utility calculus might be used to characterize a notion of a 'trust domain'. Within a system model, with an agent is represented as a process, at any given point in the agent's execution, the process is associated with a location (which we suppress for now) within the system and has access to a collection of resources. That is, the agent has a state. As described above, the agent is also associated with a utility function. Here we interpret the utility function as a loss function, associating a *cost* $k_E(a_i)$ with each choice $a_i$ that is made as a process executes, so that the trace $\sigma$ of the process that describes agent $E$ gives the total cost $K$ of an agent $E$'s execution:

$$K_E(\sigma) \;=\; \sum_{\sigma = a_1, \ldots, a_k} k_E(a_i). \qquad (11)$$

For now, we consider just finite traces.

The intended situation is depicted in Figure 6. Here the need for the concept of location should be apparent. Indeed, a logical or physical location would seem to be an essential component of the intended notion of domain. Informally, located agents manipulate their resource environments, but, in our formulation, they do so in contexts which characterize the extent to which they do so whilst maintaining a required logical property (intuitively, the 'trust' property) within a specified bound on cost. This approach stands in contrast to approaches in which constraints are expressed purely in terms of preferences, where impossible choices, that can be expressed logically in our setting, must represented by 'infinitely negative' utility. For brevity, we will not employ location explicitly, along the lines of the discussion in Section 2, instead trusting that the intuitions suggested in Figure 6 will make a sufficiently strong suggestion.

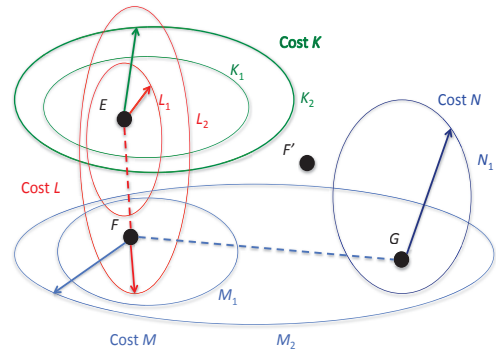

**Figure 3: Iso-utilities and Trust Domains**

Here the agent $E$ may be given one of two different choices of cost (utility) function. If $K_E = K$, then $F$ is not within $E$'s trust domain at either the $K_1$ or $K_2$ levels. If, however, $K_E = L$, then $F$ is within $E$'s trust domain at the $L_2$, but not at the $L_1$ level. Agent $F$'s cost function, $M$, includes agent $G$ at the $M_2$ level, but not at the $M_1$ level ($M_1 \leq M_2$). $F'$ is in no-one's domain at any of the given levels of utility.

The formal definition of a trust domain is set up, using the process-utility calculus, for an agent $E$ together with a property $\phi$ required (by the agent or by the designer/manager) of the part of the system or collection of agents that is to be trusted, the agent's utility function $K_E$, which assigns values to choices made as the agent executes, and a bound $K$ on the total cost of the trace, which characterizes the total acceptable cost to the agent in reaching or interacting with other parts of the system or other agents within it.

The trust domain is then constructed as a collection of contexts within which the agent may evolve whilst maintaining the properties by which it determines trust. Two properties are required to establish a viable definition. First, a bound $K$ on the cost that $E$ is prepared to incur. Second, a propositional assertion $\phi$ about the state to which $E$ can evolve within that cost constraint. So, if $R$ is the resource initially associated with $E$, then we can define, building on (11),

$$\text{TD}(E, \phi, K_E, K) = \Big\{ C \mid \text{there exist closed } F \text{ and trace}$$
$$\sigma \text{ such that } R, E \xrightarrow[C]{(e,\mathbf{1})}^{\sigma} S, F$$
$$\text{and} \quad S, F \models_{C'} \phi$$
$$\text{and} \quad K_E(\sigma) \leq K \Big\}, \quad (12)$$

where the resource $S$ is that which is derived from $R$ by the the trace $\sigma$ and $C \xrightarrow[C]{R,E}^{\sigma'} C'$, where $\sigma'$ is the trace of context actions corresponding to $\sigma$.

Notice that the inner context is empty, so imposing no restriction on the evolutions considered. More generally, in further research, we could generalize the definition to consider trust domains for agents $A$ with non-empty inner contexts, corresponding to a degree of under-specification. A richer treatment of logical satisfaction would then be needed.

Extending our banking example, we can give a simple sketch of how trust domains work. Consider a process

$$Banker \times Lawyer \times P, \quad (13)$$

where $P$ is either *Attacker* or *Client*, and where we modify *Banker* and introduce *Lawyer*, as follows:

$Banker = shareL : Banker' +_K notshareL : Banker'$
$Lawyer = 1 : (shareP : Lawyer' +_L notshareP : Lawyer')$.

In terms of Figure 6, *Banker* corresponds to $A$, *Lawyer* to $B$, and $P$ to $C$. Letting *Banker*'s cost be $L$, *Lawyer*'s cost be $M$, and letting $shareL$, $shareC$, and $shareA$, etc., be the evident sharing (of data, say) actions with lawyer, client, and attacker, we obtain

$L(shareL : Banker' \times 1 : Lawyer \times 1 : Attacker) \geq$
$L(notshareL : Banker' \times 1 : Lawyer \times 1 : Attacker)$,

but

$L(shareL : Banker' \times 1 : Lawyer \times 1 : Client) \leq$
$L(notshareL : Banker' \times 1 : Lawyer \times 1 : Client)$,

and

$M(1 : Banker' \times shareA : Lawyer \times 1 : Attacker) \geq$
$M(1 : Banker' \times notshareA : Lawyer \times 1 : Attacker)$,

but

$M(1 : Banker' \times shareC : Lawyer \times 1 : Client) \leq$
$M(1 : Banker' \times notshareC : Lawyer \times 1 : Client)$,

and see that *Banker*'s trust domain for sharing will include *Lawyer* and *Client*, but not *Attacker*.

Here, the proposition $\phi$ in (12) would be something like $\phi_{Banker}$ = 'the bank retains a good credit rating while sharing data'. Different $\phi$'s give different domains.

We remark that work in the economics tradition would tend to code propositional constraints within utility, and that work in logic would tend to code utility constraints propositionally. In our setting, the structure provided by the Hennessy–Milner-style logic suggests it is natural to maintain the distinction between utility and logical properties. Further work from this section is to consider information flow [1, 4] between trust domains.

$$\frac{\overline{R_B, present : Banker' \xrightarrow[C_C^p]{C_1} {}^{present} R_B, Banker'}}{\frac{R_B, Banker \xrightarrow[C_C]{C_1} {}^{present} R_B, Banker'}{R_C \circ R_B, Client \times Banker \xrightarrow[C_\emptyset]{C_1} {}^{logIn, present} \mu((logIn, present), R_C \circ R_B), Client' \times Banker'}} \text{(Prefix)} \quad \begin{array}{c} u_B(R, Client \times (idle_B : Banker')) \leq \\ u_B(R, Client \times (present : Banker')) \end{array} \text{(Sum)}$$
$$\text{(Prod)}$$

where $C_1 = e, \mathbf{1}$, $C_C = R_C, Client \times []$, $C_C^p = R_C, Client \times ([] +_{u_B} idle_B : Banker')$

**Figure 4: Banker choice in Client context**

$$\frac{\overline{R_B, idle_B : Banker' \xrightarrow[C_A^i]{C_1} {}^{idle_B} R_B, Banker'}}{\frac{R_B, Banker \xrightarrow[C_A]{C_1} {}^{idle_B} R_B, Banker'}{R_A \circ R_B, Attacker \times Banker \xrightarrow[C_\emptyset]{C_1} {}^{idle_A, idle_B} \mu((idle_A, idle_B), R_A \circ R_B), Attacker' \times Banker'}} \text{(Prefix)} \quad \begin{array}{c} u_B(R, Attacker \times (present : Banker')) \leq \\ u_B(R, Attacker \times (idle_B : Banker')) \end{array} \text{(Sum)}$$
$$\text{(Prod)}$$

where $C_1 = e, \mathbf{1}$, $C_A = R_A, Attacker \times []$, $C_A^i = R_A, Attacker \times (present : Banker' +_{u_B} [])$

**Figure 5: Banker choice in Attacker context**

# APPENDIX
# A. DERIVATIONS OF EXAMPLES

We provide a detailed derivation of the examples introduced in Equations 3–10. This example consists of situationally dependent choices that make use of joint access control. In order to encode the joint access control, we make use of semaphore resources. We let $R$s stand for sets of atomic resources, such as $Acnt$, $USB$, $r_i$s, etc., and make use of the $\rho$ notation [12, 11], defined by

$$\rho(a) = \min \{R \mid \mu(a, R) \downarrow\},$$

to denote which resources are required for the modification function to be defined for a given action:

$$\rho(logIn) = \{Acnt, r_1\} = R_C \qquad \rho(idle_C) = \{r_2\}$$
$$\rho(present) = \{USB, r_2\} = R_B \qquad \rho(idle_B) = \{r_1\}.$$

This ensures that the $idle_C$ and $present$ actions, and the $idle_B$ and $logIn$ actions, cannot co-occur in a reduction, as they require the same semaphore resources. We then define the modification function for each action:

$$\mu(logIn, R_C) = R_C \qquad \mu(idle_C, \{r_2\}) = \{r_2\}$$
$$\mu(present, R_B) = R_B \qquad \mu(idle_B, \{r_2\}) = \{r_2\}.$$

Let $R = \{Acnt, r_1, USB, r_2\}$. In order to denote our preference of giving the presentation over idling, in the presence of the client and the absence of the attacker, we define a portion of the banker's preference function as

$$u_B(R, Client \times (present : Banker')) = 0.7$$
$$u_B(R, Client \times (idle_B : Banker')) = 0.5.$$

We then have the reduction

$$R_C \circ R_B, Client \times Banker \xrightarrow[C_\emptyset]{C_1} {}^{logIn, present}$$
$$\mu((logIn, present), R_C \circ R_B), Client' \times Banker',$$

as derived in Figure 4. We also have the property

$$R_B, Banker \vDash_{C_C} \langle present \rangle \top.$$

This can be derived using the satisfaction relation in Figure 2, specifically the case for the diamond modality, as

$$R_B, Banker \xrightarrow[C_C]{C_1} {}^{present} \mu((present), R_B), Banker',$$

by Figure 4.

In order to encode the joint access control we make use of semaphore resources, as defined which resources are required for the attacker's actions:

$$\rho(attack) = \{r_1\} = R_A \qquad \rho(idle_A) = \{r_2\}.$$

and define the modification function for the attackers actions

$$\mu(attack, R_A) = R_A \qquad \mu(idle_A, \{r_c\}) = \{r_c\}$$

To express the banker's preference to idle in the presence of an attacker, we define a further portion of its preference function, and give a higher utility to idling in such a situation:

$$u_B(R, Attacker \times (present : Banker')) = 0.1$$
$$u_B(R, Attacker \times (idle_B : Banker')) = 0.2.$$

Here we have the reduction

$$R_A \circ R_B, Attacker \times Banker \xrightarrow[C_\emptyset]{C_1} {}^{idle_A, idle_B}$$
$$\mu((idle_A, idle_A), R_A \circ R_B), Client' \times Banker'$$

as derived in Figure 5. We also have the following property:

$$R_B, Banker \vDash_{C_A} \neg \langle present \rangle \top.$$

This can be derived using the satisfaction relation in Figure 2. By the diamond modality, as $R_B, Banker \not\xrightarrow[C_A]{C_1} {}^{present}$, by Figure 4, we have that $R_b, Banker \not\vDash_{C_A} \langle present \rangle \top$. Then the property follows directly by the interpretation of negation.